\documentclass[11pt]{article} 
\usepackage{amsmath}
\usepackage[utf8]{inputenc} 

\usepackage{geometry} 
\usepackage{graphicx} 

\usepackage{amsmath}
\usepackage{amssymb}
\usepackage{booktabs} 
\usepackage{array} 
\usepackage{paralist} 
\usepackage{verbatim} 
\usepackage{subfig} 

\usepackage{fancyhdr} 
\usepackage{sectsty}
\usepackage{hyperref}
\allsectionsfont{\sffamily\mdseries\upshape} 
\usepackage[english]{babel}
\usepackage{authblk}
\usepackage[nottoc,notlof,notlot]{tocbibind} 
\usepackage[titles,subfigure]{tocloft} 

\newcommand{\beq}{\begin{equation}}
\newcommand{\eeq}{\end{equation}}

\begin{document}

\title{Persistent current in a thin superconducting wire}

\author{Ilya Vilkoviskiy \\ email: \href{mailto:reminguk@gmail.com}{reminguk@gmail.com}}
\affil{Moscow Institute of Physics and Technology, 141700 Dolgoprudny, Russia}

\affil{I.E. Tamm Department of Theoretical Physics, P.N. Lebedev Physical
Institute,\\ Leninsky ave. 53, 119991 Moscow, Russia}

\affil{Center for Advanced Studies,
Skolkovo Institute of Science and Technology,\\
143026 Moscow, Russia}

\maketitle
\begin{abstract}{In this paper, we explore the persistent current in thin superconducting wires and accurately examine the effects of the phase slips on that current. The main result of the paper is the formula for persistent current in terms of the solutions of certain (nonlinear) integral equation. This equation allows to find asymptotics of the current at long(small) length of the wire, in that paper, we interested in the region in which the system becomes strongly interacting and very few amounts of information can be extracted by perturbation theory. Nevertheless, due to the integrability, exact results for the current can be obtained. We observe that at the limit of a long wire, the current becomes exponentially small, we believe that it is the signal that phase slips may destroy superconductivity for long wires, below BKT phase transition.}
\end{abstract}
\newpage
\tableofcontents
\newpage

\section{Introduction}

    The quantum effects in strongly interacting systems is a very interesting theme. It is known, that thin superconducting wires may lost two general features of superconductors : zero resistivity and existence of dissipationless (persistent) current in an external magnetic flux. This effects was observed at the whole range of temperatures below the critical $T_c$, even at $T \to 0$ \cite{Arutyun}. Which means that effect of superconductivity breaking, is of a quantum nature because thermal fluctuations is suppressed at small  temperatures. It was realised \cite{ZGOZ} that these effects caused by the one particular process - quantum phase slip (QPS). To understand what is the QPS, let us assume that low energy behaviour of the system could be described by the two fields : module and phase of the order parameter $\Delta(x,t)=|\Delta(x,t)| e^{i\phi(x,t)}$ , note that as $\phi$ is a phase, it is defined only modulo $2\pi$ : $\phi \sim \phi +2\pi$. If we look at the two dimensional Euclidean plane $(x,\tau=i t)$ , the phase slip could be thought as the following field $\phi$ configuration
\beq
\phi(x,\tau)=\arctan(\frac{x}{\tau})
\eeq
While $|\Delta|$ is nontrivial only in a small region near the point $x=0, \tau =0$ : $|\Delta| \sim const $ for $x^2+\tau^2 \gg 1$ . Though such solutions have a singularity at $\tau=0$ for small $\tau$ , they looks like a step function, and due to the fact that $\phi$ is a phase very close to the configurations $\phi = const$.
\begin{figure}[h]
\begin{minipage}[h]{1 \linewidth}
\center{\includegraphics[width=0.7\linewidth]{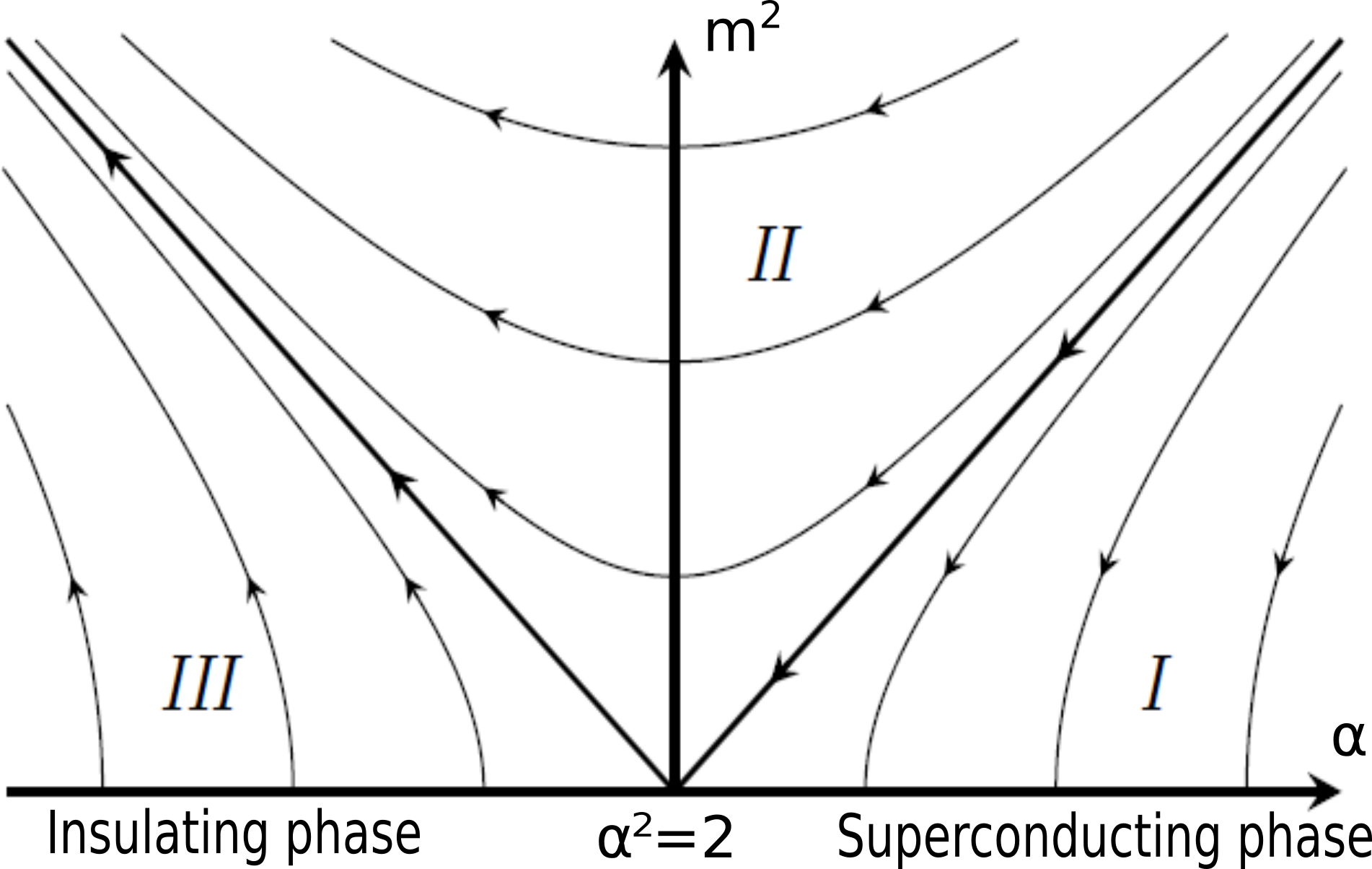}}
\end{minipage}
\caption{BKT phase diagram, $m^2$ - is a phase slips rate. $\alpha^2\sim \sqrt{S}$  , where S is the wire cross section}
\label{wire1}
\end{figure}                                                                                                                
    In this paper, we examine the phase and wire length dependence of the persistent current at zero temperature, in order to separate the quantum and thermal fluctuations. It is known that under varying the thickness of the wire the system undergoes a BKT phase transition see fig 1. The phase $\alpha^2>2 $ is a superconducting one and has been investigated in \cite{AS}, the opposite phase $ \alpha^2<2 $ , which as we will argue is insulating one, will be the subject of our interest. This phase hasn't been investigated yet, both because of lack of experiments and also because there is no simple theoretical model for that region. In that paper, we try to fill this gap. 
        
        Generally, in the superconducting phase, we have some macroscopic theory at high energies, then, if the initial coupling is small enough we can use Renormalisation group to scale to the lower energies (and larger distances) and arrive at the weakly interacting system at the limit of large distances. That means that at the large distances we can treat the system perturbatively, effects of interaction becomes stronger when one goes to the smallest distances, here, in order to calculate current one can use quantum mechanical approximation - which neglects coordinate dependence of fields: $\phi(x,t) \to \phi (t)$ \cite{AZS}. Instead in the insulating phase, system is asymptotically free at the small distances and strong interacting at large ones, physically it means that we expect that current will  have simple sawtooth-like behaviour for the rings of small circumference and experience some unknown behaviour at the limit of large (compared to the only dimensional parameter of the system - the phase slips rate) circumference. This fact makes the insulating region interesting both for theoretical and experimental physicists.
        
            It is very hard to analyze the current in the whole range of lengths because there is no any simple model for wires of a large circumference, instead, one can use underlying integrability in order to derive an exact equation for the current, which then can be solved numerically or analysed exactly in the limits of large or small wires, see appendix D for details.
            
            The paper is organized as follows: at the section 2 we discuss the physical system and make a review of the effective model for 1-dimensional superconducting rings. At the section 3 we argue what interacting terms could be added to the Lagrangian, and argue that all physics is dominated by the one term - which bring us to the Sine-Gordon theory.  At the section 3 we present the results of the paper. The ground state energy and the current could be written in terms of solutions of Destri-De Vega equation (DDV) , (\ref{DDV1}), the details explained  in the Appendix. In the Appendix A we review dualities between bosons and fermions and relations between their statistical sums (\ref{bosonisation}) .In Appendix B we sketchy explain the relations between XXZ spin chains and Thirring model. Relations between XXZ Bethe Ansatz (\ref{BAE}) and nonlinear equations (\ref{DDV}) which we finally use as the main computational tool explained in the Appendix C. Finally in the Appendix D , we show how to calculate persistent current , using the DDV equation.
\section{Physical system}
 \begin{figure}[h]
\begin{minipage}[h]{1 \linewidth}
\center{\includegraphics[width=0.5\linewidth]{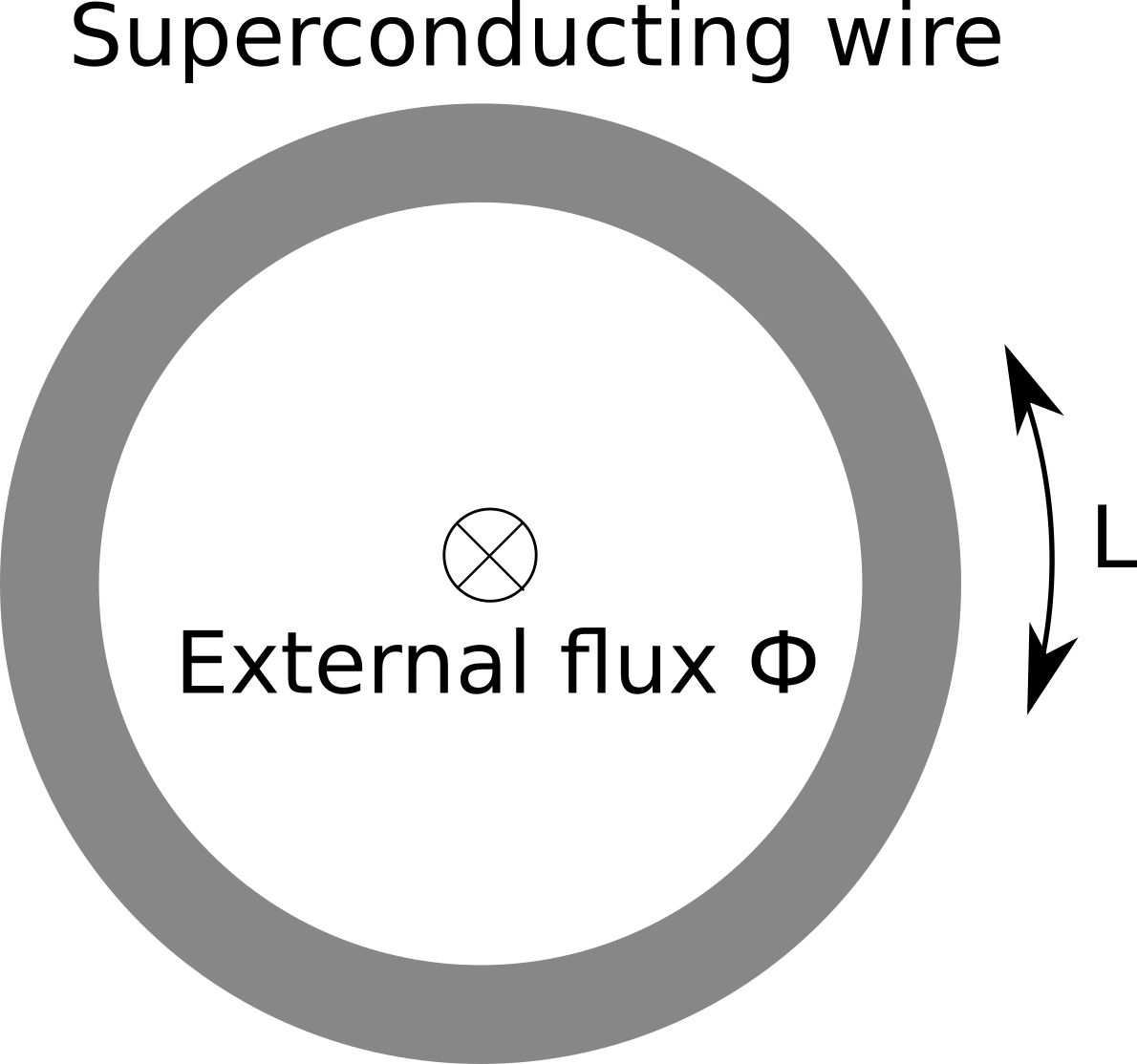} }
\end{minipage}
\caption{Superconducting ring of the circumference L, in the external magnetic flux}
\label{wire1}
\end{figure}
We consider a superconducting ring of the circumference L, at the limit of small temperature $T\to 0$. Low energy physics of the wire is described by the order parameter 
$\Delta=|\Delta(x,t)|e^{i\phi(x,t)}$ , moreover as fluctuations of the $ |\Delta(x,t)|$ described by a massive field we assume that at low energies $ |\Delta(x,t)|$ is an independent of 
$\{x,t\}$ constant. \\
Effective theory described be the Lagrangian , see \cite{AS} for details:\\
\beq
L=\frac{\alpha^2}{4
\pi } \left(\frac{1}{2}(v \partial_t \phi(x,t))^2 -\frac{1}{2}(v^{-1}\partial_x \phi(x,t))^2\right) +m^2 V_{slips} \label{MS}
\eeq

As was argued in \cite{AS} : for the standard dirty metals the only dimensionless parameter which control the physics of the system is $\frac{\alpha^2}{4\pi}=\frac{1}{2e^2}\sqrt{\frac{C_w}{L_{kin}}}$ , $v^2=\frac{\pi \sigma S}{C_w}$ where $C_w$-is the geometric wire capacitance per length and  $L_{kin}=\frac{1}{\pi \sigma_N S}$- is the kinetic wire inductance  , $\sigma_N$ -  is the normal state Drude conductance of the wire and $S$ is the wire cross section.

The term $M V_{slips}$ is introduced to take phase slips into account, and we will investigate it carefully below. 
As usual , the current is the gradient of an order parameter phase $J=\frac{\sigma}{2e}\partial_x \phi$, and the charge density $\rho=V C_w=-\frac{C}{e}\partial_t \phi$, so that the continuity equation $\partial_t \rho+\partial_x J=0$ turns to an equation of motion for the field $\phi$ 
\beq
v^2 \partial_t^2\phi -\partial_x^2\phi=0
\eeq
In order to couple the external electro-magnetic field we should analyze the transformation law of field $\phi$ under the gauge transformations
\begin{eqnarray}
A_{\mu}\to A_{\mu}+\partial_{\mu}f \\
\Delta\to \Delta e^{i f}
\end{eqnarray}
The gauge invariant Lagrangian is
\beq
L=\frac{\alpha^2}{2
\pi } \left(\frac{1}{2}(\partial_t \phi(x,t)-A_{t})^2 -\frac{1}{2}(\partial_x \phi(x,t)-A_x)^2\right) \label{L}
\eeq
In order to introduce external magnetic flux, one can choose $A_{\mu}$ in the form
\begin{eqnarray}
A_t=0 \\
A_x=\frac{\Phi}{L\alpha^2}
\end{eqnarray}
Here we choose the normalization of the flux $\Phi$ in such a way that it is $2\pi$ periodic: statistical sum is periodic function of $\Phi$ with the period $2\pi$.
\\
To the end of this section, let us note that velocity in the Lagrangian [\ref{L}] can be scaled out by redefinition of coordinate $x \to v x$ and time $t\to v t$ , later, for simplicity, we will use this convention.

\subsection{Mathematical model}
We will study the Sine-Gordon theory
\beq
S=\int\limits_0 ^{\beta \sim \infty} \int\limits_{0}^L[\frac{1}{2\pi \alpha^2} \left(\frac{1}{2}(\partial_t \chi(x,t))^2 -\frac{1}{2}(\partial_x \chi(x,t))^2+m^2 \cos(\chi(x,t))\right)- \frac{\Phi}{2\pi L} \partial_t \chi(x,t)] dx dt \label{A}
\eeq
There are several ways to arrive to this effective theory, which takes phase slips into account, in  that paper we try to write it down from general principles, but it also can be derived as an effective theory of phase slips - which are vortices in the imaginary time formalism \cite{AS}
.\\

Starting from the free theory, it is natural to ask in which ways this theory can be deformed to an interacting one and what is the physical meaning of that deformation.
\\
As $\phi$ is  a phase, it is natural to consider following operators
\beq
V_{m}(x,t)=e^{i(m \phi(x,t)} , m\in \mathbb{Z}
\eeq
This operators has fixed conformal dimension $\Delta=\frac{m^2}{2\alpha^2}$, and mutual local to each other, that means that the product
\beq
V_{m_1}(x,t)V_{m_2}(y,\tau)=((x-y)^2+(t-\tau)^2)^{\frac{m_1m_2}{2\alpha^2}}:V_{m_1}(x,t)V_{m_2}(y,\tau):
\eeq
is single valued operator in the complex plane : $z=x+it$ , $\bar{z}=x-it$, here we assume that $:V_{m_1}(x,t)V_{m_2}(y,\tau):$ means the normal ordered product, which is meromorphic in $z$, $\bar{z}$. It is possible to enlarge the algebra of local operators, introducing the dual field 
\begin{gather}
\partial_x \chi(x,t)=\partial_t \phi(x,t)\\
\partial_t \chi(x,t)=-\partial_x \phi(x,t)
\end{gather}
Consider the following operators \cite{ZZ} :
\beq
V_{m,n}(z,\bar{z})=e^{i(m \phi(z,\bar{z})+n\alpha^2\chi(z,\bar{z}))}
\eeq
\beq
V_{m_1,n_1}(z,\bar{z})V_{m_2,n_2}(0,0)=|z|^{(\frac{m_1m_2}{\alpha^2}+n_1n_2\alpha^2)}z^{\frac{m_1n_2+n_1m_2}{2}}{\bar{z}}^{-\frac{m_1n_2+n_1m_2}{2}}:V_{m_1,n_1}(z,\bar{z})V_{m_2,n_2}(0,0):
\eeq
Which are mutual local to each other at the integers n, it is imply that $\chi$ is compactifyed on a circle of radius $\frac{1}{\alpha^{2}}$ ,  $\chi\sim \chi+\frac{2\pi}{\alpha^2}$\\
We also want our Lagrangian to be invariant under the global gauge transformations: $\phi \to \phi+f, \chi\to \chi$, general Lagrangian satisfying this restrictions contains only operators $V_{0,n}$, obviously it can be written down in terms of field $\chi$ only (also it is useful to make a little redefinition $\chi\to \frac{1}{\alpha^{2}} \chi$:
\beq
L=\frac{1}{2\pi \alpha^2} \left(\frac{1}{2}(\partial_t \chi(x,t))^2+\frac{1}{2}(\partial_x \chi(x,t))^2+\sum\limits_{n}m^2_n \cos(n \chi(x,t))\right)
\eeq
Note that operators $e^{\pm i(n\chi(z,\bar{z})}$ can be identified with operators of creation/annihilation of vortexes (phase slips) with vorticity n, see \cite{AZS} for more details and physical interpretation.\\
Here the masses $m_n$ is the dimensional parameters, with the dimension
\beq
[m_n]=2(1-\frac{(n\alpha)^2}{4})
\eeq

Easy to understand that at the low energies (high distances L)the ratio $\frac{m_n}{m_1}$ scales to zero as $L^{(1-n^2)\frac{\alpha^2}{4}}$, so, it is natural to leave only first of them. This is a standard way to get a Sine-Gordon Lagrangian.
\beq
L=\frac{1}{2\pi \alpha^2} \left(\frac{1}{2}(\partial_t \chi(x,t))^2 +\frac{1}{2}(\partial_x \chi(x,t))^2+ M^2 \cos(\chi(x,t))\right)
\eeq
It is important to note that this theory have two qualitatively different regimes :$\alpha>2 , \alpha<2$ in the first regime mass M scales to zero at high distances and perturbation theory works perfectly, in the second regime mass M scales to infinity which breaks perturbation theory down. The transition between these two regimes known an BKT transition.\\
In this paper we will work at the region $ \alpha<2$ at which Sine-Gordon theory is integrable. Natural question to be interested in is how that mass parameter affect the superconductivity, the easiest one example is the computation of the ground state energy and persistent current of the Sine-Gordon theory  in the presence of external magnetic flux $\Phi$, at zero temperature $T=\frac{1}{\beta} \to 0$
\beq
S=\int\limits_0 ^{\beta \sim \infty} \int\limits_{0}^L[\frac{1}{2\pi \alpha^2} \left(\frac{1}{2}(\partial_t \chi(x,t))^2 -\frac{1}{2}(\partial_x \chi(x,t))^2+m^2 \cos(\chi(x,t))\right)- \frac{\Phi}{2\pi L} \partial_t \chi(x,t)] dx dt \label{A}
\eeq
\beq
J=\frac{L}{2\pi \beta}\frac{\partial \log(Z(\Phi))}{\partial \Phi}=\frac{L}{2\pi}\frac{\partial E_{vac}(\Phi)}{\partial \Phi}
\eeq
The best way to solve this problem was provided by Destri and De-Vega \cite{Light-Cone}.
Below, I review their results, and apply them to our problem:

\section{Results}
Our aim is to compute statistical sum, ground state energy and current at the limit of zero temperature, here and later we will work in the imaginary time formalism:
\beq
Z=\int D\phi \ \ e^{-S_{cl}[\phi]} \sim e^{-\beta E_{vac}}
\eeq
Where $S_{cl}$ is given by the formula (\ref{A})
\beq
J=\frac{L}{2\pi \beta}\frac{\partial \log(Z(\Phi))}{\partial \Phi}=\frac{L}{2\pi}\frac{\partial E_{vac}(\Phi)}{\partial \Phi}
\eeq

Classical result (at $m=0$) tells that 
\beq
E_{vac}(\Phi)=-\frac{\pi}{6L}(1-3\frac{\Phi^2 \alpha^2}{2\pi }) \label{Evac<}
\eeq
\beq
J(\Phi)=\frac{\alpha^2 \Phi}{4\pi^2} \label{Jphi}
\eeq
For the case $m^2\ne0$,  we will write down an intregral equation for a ground state energy (\ref{DDV}) , \cite{DDV}, see appendix for the details.
\beq
\epsilon(\theta)=-M L \cosh(\theta)-i\Phi-G\star\log(1+e^{\epsilon})+G_{c}\star \log(1+e^{\bar{\epsilon}}) \label{DDV1}
\eeq
Where $\star$ is a convolution, and $G_c$ ,   $G$ are known functions (\ref{G})
\beq
G(\theta)=\int\limits_{0}^{\infty} \frac{dk}{4\pi}\frac{\sinh(\pi k(\frac{\pi}{2\gamma}-1))}{\sinh(\frac{\pi k}{2} (\frac{\pi}{\gamma}-1))\cosh(\frac{\pi k}{2})}\cos(k \theta)
\eeq
\beq
G_c(\lambda)=G(\lambda+i\pi)
\eeq
\beq
E=-M \int\limits_{-\infty}^{\infty} \cosh(\theta)\left(\log(1+e^{\epsilon})+\log(1+e^{\bar {\epsilon}})\right) \frac{d\theta}{2\pi} \label{E}
\eeq

It is simple to get from this equation that in the limit of large $M L\to \infty$ 
\beq
E=-M \left(\frac{2}{\pi}\right)^{3/2}e^{-(M L-\frac{3}{4}\pi)}\cos(\Phi) +o(e^{-2M L}) \label{Evac>}
\eeq
\beq
J=\frac{M L\sqrt{2}}{\pi^{\frac{5}{2}}} e^{-(M L-\frac{3}{4}\pi)}\sin(\Phi) +o(e^{-2M L})
\eeq
Here $M $ is the soliton mass, it is related to the bare mass $m$ through the formula \cite{ZM}
\beq
 M=\left(\frac{\pi m \Gamma\left(\frac{2-2\alpha^2}{2}\right)}{\Gamma\left(\frac{\alpha^2}{2}\right)}\right)^{\alpha^2-2}\frac{2\Gamma\left(\frac{\alpha^2}{2(2-\alpha^2)}\right)}{\sqrt{\pi}\Gamma\left(\frac{1}{2-\alpha^2}\right)}
\eeq
Higher corrections could be also computed order by order. The technique allows one to compute the current numerically, here is the plots of the current $J({\Phi})$, for different values of $M L$
 \begin{figure}[h]
\begin{minipage}[h]{1 \linewidth}
\center{\includegraphics[width=0.75\linewidth]{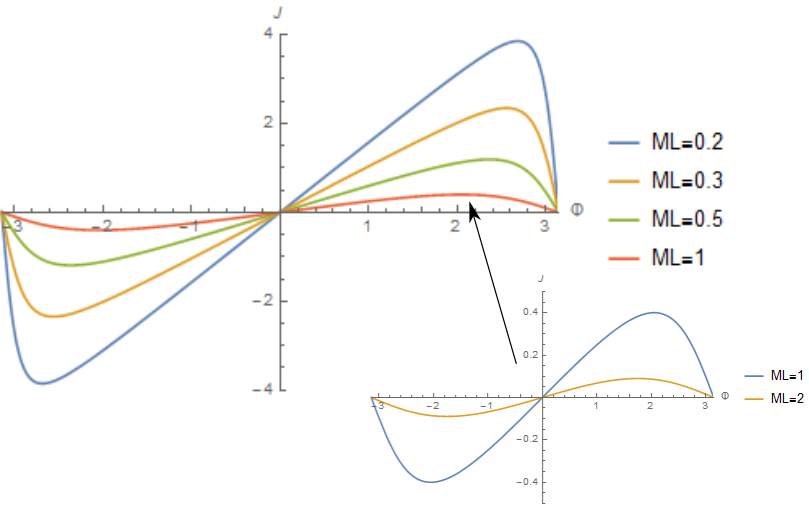} }
\end{minipage}
\caption{For different values of $ML$ , Current $J({\Phi})$ - interpolates between Sawtooth and Sine}
\label{ris:image1}
\end{figure}
\newpage
\section{Discussion}
In this paper we started investigation of the so-called `insulating' region of thin superconducting nanowires. Following the paper of A.Semenov and A.Zaikin we used Sine-Gordon theory as an effective low energy theory for our system. In order to explore superconducting properties of the ring we computed the simplest quantity of superconductors : the persistent current dependence on the external magnetic flux and length of the ring. We found that for the ring of large lengths the current is exponentially suppresed $J\sim e^{-M  L}$. This result can not be obtained in pertrubation theory, instead we used integrability of the theory in order to find asymptotics of the current exactly. Our results agrees with the results of $\cite{AS}$ , which was obtained with the help of renorm group argument, which are not very confident at the regime of strong coupling. Although, our exact computation shows that heuristic argument of renorm group leads to a qualitatively correct results. Also our method reproduce exponential and pre-exponential factors of current dependence as well as systematical computation of higher order of pertrubation series in ($e^{-ML})$.
    
    Though the current is suppressed in the bulk of superconductor, we can't confidently claim that superconductivity is broken, for example, it would be interesting to compute resistivity of the wire connected to the resistor, which is yet a nontrivial problem.
    We believe that it will be very interesting to discover the region near, and below, the BKT phase transition experimentally. Not only because it may have applications in nanoelectronics, but because it can boost our understanding of Sine-Gordon theory.  
\centerline{\bf Acknowledgements}
I thank Andrew G. Semenov for introducing me to the theme, and for interesting discussions during the work. I also thank Andey Marshakov for permanent support and his interest to this study. Finally I would like to thank Andrei D. Zaikin for his comments on earlier versions of the paper.

\newpage
\appendix
\section{Free field theory and boson-fermion correspondence, from Sine-Gordon to Thirring}
In this section, we rigorously show that the theory of one bosonic field $\phi(x,t)$ - compactified on a circle : $\phi(x,t) \sim \phi(x,t)+2\pi $ is equivalent to a theory of Dirac fermions. \\
Consider , for the moment pure Sine Gordon theory:
\beq
S=\int\limits_0 ^{\beta} \int\limits_{0}^L\frac{1}{2\pi \alpha^2} \left(\frac{1}{2}(\partial_t \chi(x,t))^2 -\frac{1}{2}(\partial_x \chi(x,t))^2+m^2 \cos(\chi(x,t))\right) dx dt 
\eeq
Field $\phi$ defined to take values in $\mathbb{R}$ ,  but (at least at the classical level) we should specify a boundary conditions, for the applications to a superconductors we should consider all fields of the form : 
\begin{gather}
\chi(x+L,t)=\chi(x,t)+2\pi n \label{con1} \\
\chi(x,t+\beta)=\chi(x,t)+2\pi m \label{con2}
\end{gather}
The best way to deal with different sectors is to introduce chemical potentials $\Phi $ , and $\mu$ in the action
\begin{gather}
S_{m,n}=\int\limits_0 ^{\beta} \int\limits_{0}^L\frac{1}{2\pi \alpha^2} \left(\frac{1}{2}(\partial_t \chi(x,t))^2 -\frac{1}{2}(\partial_x \chi(x,t))^2+m^2 \cos(\chi(x,t))\right) dx dt+m\Phi +n\mu 
\end{gather}
\beq
Z(\Phi,\mu)=\sum\limits_{m,n \in \mathbb{Z}} Z_{m,n}=\sum\limits_{m,n \in \mathbb{Z}} \int_{m,n} D\phi e^{iS_{m,n}}
\eeq
Where integration $\int_{m,n}$ performed under the fields satisfying conditions (\ref{con1},\ref{con2}).
It is easy to understand that
\beq
Z(\Phi,\mu)=\int D\phi e^{i S} \   \ e^{i\frac{\Phi}{2\pi L}  \int\limits_0 ^{\beta } \int\limits_{0}^L\partial_t \chi(x,t) dx dt+i\frac{\mu}{2\pi \beta}  \int\limits_0 ^{\beta } \int\limits_{0}^L\partial_x \chi(x,t) dx dt }
\eeq
Where integration performed in all sectors. Alternatively
\beq
Z_{m,n}=\int\limits_{0}^{2\pi}\int\limits_{0}^{2\pi} Z(\Phi,\mu)e^{i n\phi +i m\mu}d\mu d\phi
\eeq
We state that the same statistical sum $Z(\Phi,\mu)$ can be obtained from the dual fermionic model (Thirring model) with the fields $\psi(x,t)=\begin{pmatrix} \psi_L(x,t) \\ \psi_R(x,t) \end{pmatrix}$ and Hamiltonian
\beq
H_{Th}=\int\left(\psi^{\dagger} \sigma^3 (-i \partial_x+\frac{\Phi}{L}) \psi+\frac{\mu}{ \beta} \psi^{\dagger}\psi+M \psi^{\dagger} \sigma^2 \psi +2g\psi^{\dagger}_R\psi_R\psi^{\dagger}_L\psi_L  \right)dx \label{Th}
\eeq

This statement can be easily proved by direct computation at the free (and massles) fermions point $g=M=0$, and than one can follow the Coleman's standard argument to understand it holds for any orders of  $g$ and $M$ in perturbation theory.
\subsection{Free boson}
\beq
S=\int\limits_0 ^{\beta} \int\limits_{0}^L\frac{1}{2\pi \alpha^2} \left(\frac{1}{2}(\partial_t \chi(x,t))^2 -\frac{1}{2}(\partial_x \chi(x,t))^2+\frac{\Phi}{L} \partial_t \chi(x,t) +\frac{\mu}{\beta}  \partial_x \chi(x,t) \right) dx dt 
\eeq
Classical solutions have the form
\beq
\chi(x,t)=\frac{2\pi n x}{L}+\frac{2\pi m t}{\beta}+\sum\limits_{k \in  \mathbb{Z}/\{0\}}\chi_k(t)e^{\frac{2\pi i k x}{L}}
\eeq
$\phi_k(t)$ just the oscillators, each of them make a contribution equal to a $\frac{1}{e^{\pi i k \frac{\beta}{L}}-e^{-\pi i k \frac{\beta}{L}}}$ , and we will get
\beq
Z_{\alpha}(\Phi,\mu)=\prod\limits_{k\ne 0}\frac{1}{e^{\pi i k \frac{\beta}{L}}-e^{-\pi i k \frac{\beta}{L}}}\sum_{m,n}e^{i \pi (\frac{\beta}{L} \frac{M^2}{\alpha^2}-\frac{L}{\beta} \frac{n^2}{\alpha^2})+i\Phi n+i\mu m}
\eeq
Or, after Poisson resummation

\beq
Z_{\alpha}(\Phi,\mu)=\prod\limits_{p\ne 0}\frac{1}{e^{\pi i p \frac{\beta}{L}}-e^{-\pi i p \frac{\beta}{L}}}\sum_{m,n}e^{i \pi (\frac{\beta}{\alpha^2 L} M^2+\frac{\alpha^2\beta}{L}( n+\frac{\Phi}{2\pi})^2)+i\mu m} \label{FB}
\eeq
At the limit of zero temperature $ (\beta\to i\frac{1}{T}\to i\infty)$
\beq
Z\sim e^{-\beta E_{vac}(\Phi)}
\eeq
\beq
E_{vac}(\Phi)=-\frac{\pi}{6L}(1-3\frac{\Phi^2 \alpha^2}{2\pi }) \label{Evac}
\eeq
Which reproduce classical answer for the current of the wire, in the presence of external flux
\beq
J(\Phi)=\frac{\alpha^2 \Phi}{4\pi^2} \label{Jphi}
\eeq
\subsection{Free fermions}
The method of Destri and De-Vega is applied to Thirring model, so it is worth to show directly that , at least at the free level, fermionic theory reproduce the same answer as bosonic.\\
We begin with the free and massles theory of Dirac fermions, and show that it's statistical sum is identical to bosonic one at the point $\alpha=\sqrt{2}$
\beq
H_{ff}=\int\left(\psi^{\dagger} \sigma^3 (-i \partial_x+\frac{\Phi}{L}) \psi+\frac{\mu}{ \beta} \psi^{\dagger}\psi \right)dx 
\eeq
\beq
\psi_{\alpha,r}=c_{\alpha,r}e^{\frac{2\pi i}{L} r}
\eeq
\beq
H_{ff}=\sum\limits_{r\in \mathbb{Z}+1/2}\frac{2\pi}{L}c^{\dagger}_{R,-r}c_{R,r}(r+\frac{\Phi}{2\pi}+\frac{\mu L}{2\pi \beta})+\sum\limits_{s\in \mathbb{Z}+1/2}\frac{2\pi}{L} c_{L,s}^{\dagger} c_{L,-s}
(s-\frac{\Phi}{2\pi}+\frac{\mu L}{2\pi \beta})
\eeq
\beq
Z_f(\Phi,\mu)=tr\left(e^{i\beta H_{ff}}\right)
\eeq

\begin{gather}
Z_f(\Phi,\mu)=\prod\limits_{p\ne 0}\frac{1}{e^{\pi i p \frac{\beta}{L}}-e^{-\pi i p \frac{\beta}{L}}}(\sum_{m\in 2\mathbb{Z},n \in \mathbb{Z}}e^{i \pi (\frac{\beta}{2 L} m^2+\frac{2\beta}{L}( n+\frac{\Phi}{2\pi})^2)+i\mu m}+ \nonumber \\+ \sum_{m\in 2\mathbb{Z}+1,n \in \mathbb{Z}+1/2}e^{i \pi (\frac{\beta}{2 L} m^2+\frac{2\beta}{L}( n+\frac{\Phi}{2\pi})^2)+i\mu m})
\end{gather}
If we in addition project to a even m (fermion charge) we obtain
\beq
Z_{f}^{even}(\Phi,\mu)=\prod\limits_{p\ne 0}\frac{1}{e^{\pi i p \frac{\beta}{L}}-e^{-\pi i p \frac{\beta}{L}}}\sum_{m\in 2\mathbb{Z},n \in \mathbb{Z}}e^{i \pi (\frac{\beta}{2 L} m^2+\frac{2\beta}{L}( n+\frac{\Phi}{2\pi})^2)+i\mu m}
\eeq
analogically
\beq
Z_{f}^{odd}(\Phi+\pi,\mu)=\prod\limits_{p\ne 0}\frac{1}{e^{\pi i p \frac{\beta}{L}}-e^{-\pi i p \frac{\beta}{L}}}\sum_{m\in 2\mathbb{Z}+1,n \in \mathbb{Z}}e^{i \pi (\frac{\beta}{2 L} m^2+\frac{2\beta}{L}( n+\frac{\Phi}{2\pi})^2)+i\mu m}
\eeq
Finally, we obtain
\beq
Z^{boson}_{\alpha=\frac{1}{\sqrt{2}}}(\Phi,\mu)=Z_{f}^{even}(\Phi,\mu)+Z_{f}^{odd}(\Phi+\pi,\mu) \label{bosonisation}
\eeq
Which simply means that we have to impose antiperiodic boundary conditions in even charge sector and periodic in odd charge sector. Of course, as the bosonic and fermionic statistical sums are identical, fermionic models reproduce the same answers for the ground state energy and current as the bosonic ones (\ref{Evac},\ref{Jphi}).
\subsection{General $\alpha$ and interacting fermions}

We assme that for a general $\alpha$ ,  formula analogical to (\ref{bosonisation}) holds
\beq
Z^{boson}_{\alpha}(\Phi,\mu)=Z_{Th,g}^{even}(\Phi,\mu)+Z_{Th,g}^{odd}(\Phi+\pi,\mu)
\eeq
Where $Z_{Th,g}^{even/odd}$ is the statistical sum of the Thirring Hamiltonian with the  antiperiodic/periodic boundary conditions in even/odd fermionic charge sector
\beq
H_{Th,g}=\int\left(\psi^{\dagger} \sigma^3 (-i \partial_x+\frac{\Phi}{L}) \psi+\frac{\mu}{ \beta} \psi^{\dagger}\psi+2g\psi^{\dagger}_R\psi_R\psi^{\dagger}_L\psi_L  \right)dx 
\eeq
\section{From Thirring to XXZ spin chain}
This section based on the work of Destri and De-Vega \cite{Light-Cone}, see also \cite{LI}\\
Here I just briefly review their results: \\
1) Sine-Gordon can be realised as a certain limit of XXZ spin chain with a number of inhomogeneities\\
2) Spectrum of energy can be analysed with the help of Bethe-Ansatz equations (BAE)\\
\beq
\prod\limits_{n=1}^N\frac{\sinh(\lambda+i\theta_n+i\gamma/2)}{\sinh(\lambda+i\theta_n-i\gamma/2)}=-e^{-2 i \Phi}\prod\limits_{j=1}^{N}\frac{\sinh(\lambda-\lambda_j+i\gamma)}{\sinh(\lambda-\lambda_j-i\gamma)}
\eeq
With the limit $\theta_i =(-1)^{i-1}\Theta/2 , \ \frac{\pi}{\gamma}\Theta=\log(\frac{2N}{M L}) \to \infty $ as $N    \to \infty $, and $\gamma$ relates to the Sine-Gordon coupling constant as \\ $\gamma=\pi(1-\alpha^2) \in [0,\pi]$ , the external magnetic flux corresponds to a twist $e^{-2 i \Phi}$ in the BAE .\\
The energy and momentum of the $\{\lambda_j \}$, configuration given by the formula:
\beq
e^{-i\frac{E\pm P}{2N}}=\prod\limits_{n=1}^N\frac{\sinh(\pm\lambda_j+\Theta/2+i\gamma/2)}{\sinh(\mp \lambda_j-\Theta/2-i\gamma/2)}
\eeq
3) This XXZ regularisation of Sine-Gordon gives the answer for the ground state energy in the presence of external magnetic flux, which (in the massless limit) agrees with the conformal one (instead of coordinate Bethe Ansatz and cut-off regularisation).
\section{From XXZ Bethe Ansatz to DDV equation}
Here we review connection between the XXZ Bethe Ansatz and nonlinear integral equation. \cite{DDV}. \\
Our starting point will be the Bethe-Ansatz equations for XXZ model with inhomogeneities
\beq
\prod\limits_{n=1}^N\frac{\sinh(\lambda+i\theta_n+i\gamma/2)}{\sinh(\lambda+i\theta_n-i\gamma/2)}=-e^{-2 i \Phi}\prod\limits_{j=1}^{N}\frac{\sinh(\lambda-\lambda_j+i\gamma)}{\sinh(\lambda-\lambda_j-i\gamma)} \label{BAE}
\eeq

Taking the logarithm of both sides , and introducing new functions
\beq
Z(\lambda)=\log\left(-e^{-2 i \Phi}\prod\limits_{j=1}^{N}\frac{\sinh(\lambda-\lambda_j+i\gamma)}{\sinh(\lambda-\lambda_j-i\gamma)}\right)-\log \left(\prod\limits_{n=1}^N\frac{\sinh(\lambda+i\theta_n+i\gamma/2)}{\sinh(\lambda+i\theta_n-i\gamma/2)}\right)
\eeq
\beq
\phi(\lambda,x)=i\log\left(\frac{\sinh(ix+\lambda)}{\sinh(ix-\lambda)}\right)
\eeq
We arrive to the following equations
\beq
Z(\lambda)=N[\phi(\lambda+\Theta,\gamma/2)+\phi(\lambda -\Theta,\gamma/2)]-\sum\limits_{j=1}^{N/2}\phi(\lambda-\lambda_k,\gamma)-2\Phi
\eeq
As usual vacuum state characterises by the condition
\beq
Z_{vac}(\lambda_j)=(-N-1+2j)\pi
\eeq
Using it, we can rewrite the sum over roots -$\lambda_k$ as a contour integral
\beq
\sum\limits_{j=1}^{N/2}\phi(\lambda-\lambda_k,\gamma)=\oint\limits_{\Gamma}  \phi(\lambda-\mu,\gamma) \frac{d}{d\mu}\log(1+e^{iZ(\mu)}) \frac{d\mu}{2\pi i} \label{sum}
\eeq
Where contour $\Gamma$ - encircles all roots $\lambda_k$, this is the way how nonlinear integral equation appears
\beq
Z(\lambda)=N[\phi(\lambda+\Theta,\gamma/2)+\phi(\lambda -\Theta,\gamma/2)]-2\Phi-\oint\limits_{\Gamma}  \phi(\lambda-\mu,\gamma) \frac{d}{d\mu}\log(1+e^{iZ(\mu)}) \frac{d\mu}{2\pi i} \label{int}
\eeq
After some manipulations \cite{DDV} this equation could be transformed to a more convenient form:
\begin{gather}
Z(\lambda)=ML \sinh(\frac{\pi \lambda}{\gamma})-\frac{\pi \Phi}{\pi -\gamma}-i  \int\limits_{-\infty}^{\infty} G(\lambda-\mu-i\eta_+) \log(1+e^{iZ(\mu+i\eta_+)}) d\mu+
\nonumber \\ 
+i  \int\limits_{-\infty}^{\infty}G(\lambda-\mu+i\eta_-,\gamma) \log(1+e^{-iZ(\mu-i\eta_-)}) d\mu
\end{gather}
\beq
G(\mu)=\frac{K}{1+K}=\int\limits_{0}^{\infty} \frac{dk}{4\pi}\frac{\sinh(k(\frac{\pi}{2}-\gamma))}{\sinh(k\frac{\pi-\gamma}{2})\cosh(\frac{\gamma k}{2})}\cos(k \mu)
\eeq
This equation (or its simple modification) is called Destri-De Vega (DDV) equation. 

\section{From DDV to asymptotics of the current in the limit of big(small) radius of the wire}
For now we will restrict ourself to a repulsive regime only ($\gamma<\frac{\pi}{2}$ and no breathers).\\ For this case Destri and De Vega suggest a prescription $\eta_+=\eta_-=\eta=\frac{1}{2}min(\gamma ; \pi -\gamma)$.  and the change of an argument : $Z(\frac{\gamma\lambda}{\pi}+i\eta)=i\epsilon(\lambda)$ , $G(\lambda) \to \frac{\gamma}{\pi}G(\frac{\gamma}{\pi}\lambda)$ we will get
\beq
G(\theta)=\int\limits_{0}^{\infty} \frac{dk}{4\pi}\frac{\sinh(\pi k(\frac{\pi}{2\gamma}-1))}{\sinh(\frac{\pi k}{2} (\frac{\pi}{\gamma}-1))\cosh(\frac{\pi k}{2})}\cos(k \theta) \label{G}
\eeq
\beq
\epsilon(\theta)=-ML \cosh(\theta)-i\Phi-G\star\log(1+e^{\epsilon})+G_{c}\star \log(1+e^{\bar{\epsilon}}) \label{DDV}
\eeq
Where $\star$ is a convolution, and $G_c$ is crossed $G$ : $G_c(\lambda)=G(\lambda+i\pi)$ , remind that energy is eqal to
\beq
E=-M\int\limits_{-\infty}^{\infty} \cosh(\theta)\left(\log(1+e^{\epsilon})+\log(1+e^{\bar {\epsilon}})\right) \frac{d\theta}{2\pi} \label{E}
\eeq
    
    Surprisingly, G coincides with the logarithmic derivative of soliton-soliton scattering matrix $G(\theta)=\frac{1}{2\pi i}\partial_{\theta}\log(S(\theta))$, and $G_c$ coincides with matrix of crossed process. So DDV equation looks similar to a TBA equation, despite the fact that it is much more simple, because in general TBA for Sine-Gordon is a system of a number of coupled nonlinear equations. 
\subsection{Large    $ML$ limit}
    Recall that equations \ref{DDV}, \ref{E} contain full information about the energy of the ground state of Sine-Gordon (on finite ring with length $L$) in the external magnetic flux $\Phi$.\\
Having this equation it is easily to analyze the large $ML$ limit, namely we can solve this equation recursively
\beq
\epsilon_0=-ML \cosh(\lambda)-i\Phi
\eeq
\beq
\epsilon_1=G\star\log(1+e^{\epsilon_0})+G_{c}\star \log(1+e^{\bar{\epsilon_0}})
\eeq
\beq
... \nonumber
\eeq
Let us make just zero order approximation, 
\beq
E=-M \int\limits_{-\infty}^{\infty} \cosh(\theta) Re\left(\log(1+e^{-ML \cosh(\frac{\pi \lambda}{\gamma})-i\Phi})\right) \frac{d\theta}{2\pi}
\eeq
Expanding logarithm in Taylor series 
\beq
E=-\frac{2M}{\pi}\sum\limits_{n \ge 1} \frac{(-1)^{n-1}}{n}K_1(n ML)\cos(n \Phi)
\eeq
For the free fermions point $(\gamma=\frac{\pi}{2})$, nonlinear part just vanishes, and it becomes an exact answer, for example we can make a conformal limit $ML\to 0$, and reproduce the classical answer (\ref{Evac},\ref{Jphi})
\beq
E=-\frac{2}{L\pi}\sum\limits_{n \ge 1} \frac{(-1)^{n-1}}{n^2}\cos(n \Phi)=-\frac{\pi}{6 L}\left(1-\frac{3}{\pi^2 } \Phi^2\right)
\eeq

Away from the free fermions point  , we can leave only the firs term from the infinite sum
\beq
E=-\frac{2M}{\pi} K_1(M L)\cos(\Phi) +o(e^{-2mL})
\eeq
\subsection{Small $ML$ limit (conformal regime)}
The analisys of small $ML$ regime is more complicated, it was done in \cite{DDV}, with the results in agreement with the conformal ones (\ref{Evac},\ref{Jphi}).
\beq
E=-\frac{\pi}{6 L}\left(1-\frac{3(\pi-\gamma)}{2\pi^2 } \Phi^2\right)+...
\eeq

\end{document}